\input epsf
\input harvmac


\newcount\figno
\figno=0

\def\fig#1#2#3{
\par\begingroup\parindent=0pt\leftskip=1cm\rightskip=1cm\parindent=0pt
\baselineskip=11pt
\global\advance\figno by 1
\midinsert
\epsfxsize=#3
\centerline{\epsfbox{#2}}
\vskip 12pt
{\bf Fig. \the\figno:} #1\par
\endinsert\endgroup\par
}

\def\figlabel#1{\xdef#1{\the\figno}}
\def\encadremath#1{\vbox{\hrule\hbox{\vrule\kern8pt\vbox{\kern8pt
\hbox{$\displaystyle #1$}\kern8pt}
\kern8pt\vrule}\hrule}}

\overfullrule=0pt

%

\def\bar{\overline}

\def\np#1#2#3{Nucl. Phys. {\bf B#1} (#2) #3}

\font\zfont = cmss10 
\font\litfont = cmr6

\def\bigone{\hbox{1\kern -.23em {\rm l}}}
\def\ZZ{\hbox{\zfont Z\kern-.4emZ}}
\def\half{{\litfont {1 \over 2}}}

\def\Re{{\rm Re ~}}
\def\Im{{\rm Im ~}}

\def\vare{\varepsilon}
%
%

\def\goto{\rightarrow}

\def\bz{\bar{z}}

\def\Kahler{K\"{a}hler}


\def\Title#1#2{\rightline{#1}
\ifx\answ\bigans\nopagenumbers\pageno0\vskip1in%
\baselineskip 15pt plus 1pt minus 1pt
\else
\def\listrefs{\footatend\vskip 1in\immediate\closeout\rfile\writestoppt
\baselineskip=14pt\centerline{{\bf References}}\bigskip{\frenchspacing%
\parindent=20pt\escapechar=` \input
refs.tmp\vfill\eject}\nonfrenchspacing}
\pageno1\vskip.8in\fi \centerline{\titlefont #2}\vskip .5in}


\Title{
\vbox{\baselineskip12pt\hbox{\tt hep-th/9712050}
\hbox{PUPT-1747}
}}
{\vbox{\centerline{String Network from M-theory}}}
\centerline{Morten Krogh\foot{krogh@princeton.edu} and 
Sangmin Lee\foot{sangmin@princeton.edu.}}
\smallskip
\centerline{\sl Joseph Henry Laboratories}
\centerline{\sl Princeton University}
\centerline{\sl Princeton, NJ 08544, USA}
\bigskip

\bigskip
\centerline{\bf Abstract}
\medskip
\noindent
We study the three string junctions and string networks in Type IIB string 
theory by explicity constructing the holomorphic embeddings of the M-theory 
membrane that describe such configurations. The main feature of them such as
supersymmetry, charge conservation and balance of tensions are derived in a 
more unified manner. We calculate the energy of the string junction and show 
that there is no binding energy associated with the junction.

\Date{December, 1997}


\nref\aha
{O. Aharony, J. Sonnenschein and S. Yankielowicz, 
\sl Interactions of Strings and D-branes from M-theory, 
\tt hep-th/9603009, \rm \np{474}{1996}{309}.}

\nref\sch
{J.H. Schwarz, \sl Lectures on Superstring and M theory Dualities, 
\tt hep-th/9607201, \rm Nucl. Phys. Proc. Suppl. {\bf 55B} (1997) 1.}

\nref\zwiebach
{M. Gaberdiel and B. Zwiebach, \sl Exceptional groups from open strings, 
\tt hep-th/9709013. \rm}

\nref\mukhi
{K. Dasgupta and S. Mukhi, \sl BPS Nature of 3-String Junctions,
\tt hep-th/9711094. \rm}

\nref\sen
{A. Sen, \sl String Network, 
\tt hep-th/9711130. \rm}

\nref\rey
{S.-J. Rey and J.-T. Yee, \sl BPS Dynamics of Triple (p,q) String Junction,
\tt hep-th/9711202. \rm}

\nref\ggt
{J. P. Gauntlett, J. Gomis, P. K. Townsend, \sl BPS Bounds for Worldvolume 
Branes, \tt hep-th/9711205. \rm}

\nref\ahaha
{O. Aharony, A. Hanany,
\sl Branes, Superpotentials and Superconformal Fixed Points,
\tt hep-th/9704170, \rm \np{504}{1997}{239}.}

\nref\kol
{B. Kol, \sl 5d Field Theories and M Theory,
\tt hep-th/9705031. \rm}

\nref\bist
{A. Brandhuber, N. Itzhaki, J. Sonnenschein, S. Theisen,
\sl On the M-Theory approach to (Compactified) 5D Field Theories,
\tt hep-th/9709010. \rm}

\nref\ahk
{O. Aharony, A. Hanany, B. Kol,
\sl Webs of (p,q) 5-branes, Five Dimensional Field Theories and Grid Diagrams,
\tt hep-th/9710116. \rm}

\nref\calmal
{C. Callan and J. Maldacena, \sl Brane Dynamics from the Born-Infeld Action, 
\tt hep-th/9708147. \rm}

\nref\gibbons
{G. Gibbons, \sl Born-Infeld Particles and Dirichlet p-Branes, 
\tt hep-th/9709027. \rm}

\nref\lpth
{S. Lee, A. Peet and L. Thorlacius, \sl Brane-Waves and Strings, 
\tt hep-th/9710097. \rm}

\nref\larus
{L. Thorlacius, \sl Born-Infeld String as a Boundary Conformal Field Theory, 
\tt hep-th/9710181. \rm}

\nref\aki
{A. Hashimoto, \sl The Shape of Branes Pulled by Strings, 
\tt hep-th/9711097. \rm}

\nref\emp
{R. Emparan, \sl Born-Infeld Strings Tunneling to D-branes, 
\tt hep-th/9711106. \rm}

\nref\BBS
{K. Becker, M. Becker and A. Strominger, \sl Fivebranes, Membranes
and Non-Perturbative String Theory, 
\tt hep-th/9507158, \rm \np{456}{1995}{130}.}

\nref\townsend
{P.K. Townsend , \sl Three Lectures on Supermembranes, 
 \rm Proceedings of the Trieste Spring School, 11.-19. April 1988.}


\newsec{Introduction}

Recently there has been a lot of interest in three string junctions in 
Type IIB string theory (IIB) \refs{\aha - \ggt }. A three string 
junction is a configuration where three strings of different type $(p,q)$
meet as shown in the figure. The configuration is planar.

\medskip
\fig{The 3-point junction}{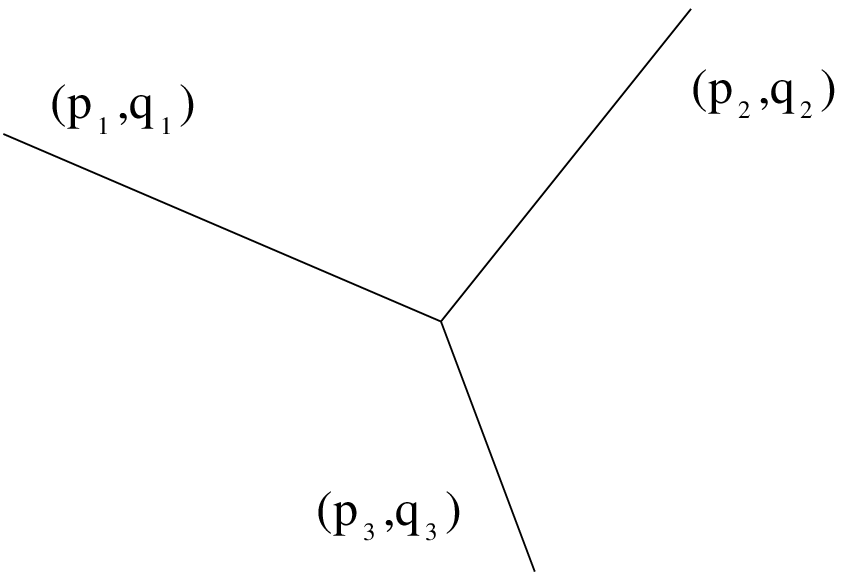}{150pt}
\figlabel\junction
\medskip

The BPS nature of this configuration was conjectured in \sch\ and proven in 
\refs{\mukhi, \sen}. It was shown that if the strings are of type 
$(p_i,q_i)$, $i=1,2,3$ then charge conservation requires 
\eqn\chargecons{
\sum_{i=1}^3 p_i = \sum_{i=1}^3 q_i = 0.
}
Furthermore the angles between the strings is determined purely by their type.
The angles are such that the total force on the vertex is zero. The tensions 
of the 3 strings counted with direction add up to zero. In \mukhi\ these 
results were derived by considerations of the Super-Yang-Mills (SYM) theory 
on a D-string. In \sen\ more general configurations, networks or 
lattice of strings, were considered. All these configurations were argued to 
be BPS configurations preserving one fourth of the supersymmetry(SUSY).

In this paper we will derive all these results by lifting the picture to 
M-theory. IIB can be viewed as M-theory on a torus in the limit where 
the torus shrinks. The strings in IIB are the membrane of M-theory 
with one direction wrapped on the torus. The $(p,q)$ type of the string
is determined by which homology cycle of the torus the membrane wraps. 
We find a smooth configuration of the 
membrane which corresponds to the three string junction in type IIB. 
The singularity at the vertex in the IIB description is removed by going 
to M-theory. In the M-theory description there is no special point.

Essentially the same technique has been used extensively to analyze five 
dimensional field theories by lifting webs of $(p,q)$ 5-branes of IIB to 
M-theory \refs{\ahaha - \ahk}. The holomorphic curves in those works 
appear as the low energy solution of the gauge theories.

The organization of the paper is as follows. 
In section 2, we derive the criterion for a membrane to be a BPS state. 
The result is that to preserve some supersymmetry the membrane has to be 
embedded as a holomorphic curve. This result is well known, but since it is 
usually not explained we include a proof of it. 
 The result is true not only for membranes in 
flat Minkowski space but also in more general settings like Calabi-Yau 
compactifications. 
In section 3, we explain the relation between $(p,q)$ strings and membranes 
of M-theory. We especially see that the orientation of the $(p,q)$ string in 
spacetime is correlated with its type in agreement with \mukhi. 
In section 4, we derive the equation for the membrane configuration 
corresponding to the three string junction, thereby giving an alternative 
proof for the BPS nature of the three string junction. The method of this 
section is generalized to construct string networks in section 5. 
In section 6, we calculate the energy of the membrane. It is seen to be 
exactly equal to the sum of the energies of each string, in accordance with
the previous analysis of fundamental strings ending on D-p-branes ($p\ge3$)
using Born-Infeld action \refs{\calmal - \emp}.
We briefly discuss possible applications of our construction in section 7.   

\newsec{Supersymmetric embedding of membranes}

In this section we will derive the condition for a membrane configuration in 
M-theory to preserve some supersymmetry. We are interested in static 
configurations, in other words the time axis on the worldvolume is directed 
along the time axis of 11 dimensional spacetime and the 2 spatial directions on 
the worldvolume are embedded in the 10 spatial dimensions of spacetime. 
We want to figure out which embeddings give rise to BPS states.
 To do that we closely follow the discussion of \BBS,
where a similar question was considered with the difference that the membrane 
was embedded as an Euclidean instanton.

The action for a membrane in M-theory is \townsend\
\eqn\memaction{
S= T_{2} \int d^3 \sigma \sqrt{-h} ( \half  h^{\alpha \beta} \partial_{\alpha}X^M
\partial_{\beta}X^N G_{MN} - \half  - i \bar{\Theta} \partial^{\alpha}X^M \Gamma_{M}
\nabla_{\alpha}\Theta + . . .)
}
Here $X^M(\sigma)$, $M=0,...,10$ describes the membrane configuration. 
$\Theta$ is an 11 dimensional Majorana spinor. 
$h_{\alpha\beta}$, $\alpha , \beta =0,1,2$ is an auxiliary worldvolume metric.
The dots denote terms of higher power in the fermi fields. The 3-form of 
M-theory has been set to zero. $G_{MN}$ is the metric of spacetime. $\Gamma_M$
are gamma matrices satisfying $\{ \Gamma_M , \Gamma_N \} = 2 G_{MN}$.
In the applications in this paper we are solely interested in the case 
$G_{MN}=\eta_{MN}$, but for the time being we can be more general and take 
spacetime to be of the form ${\bf R}^{1,0} \times K^{10}$, where 
${\bf R}^{1,0}$ is time and $K^{10}$ is a \Kahler\ manifold. 
This would cover both flat spacetime and Calabi-Yau compactifications.

The equation of motion for $h_{\alpha \beta}$ sets it equal to the induced 
metric
\eqn\fmetric{
h_{\alpha \beta} = \partial_{\alpha} X^M \partial_{\beta}X^N G_{MN}
}
The action has two fermionic symmetries. One is the global SUSY transformation
\eqn\glosym{\eqalign{
\delta_{\varepsilon} \Theta & = \varepsilon  \cr
\delta_{\varepsilon} X^M & = i \bar{\varepsilon} \Gamma^M \Theta
}}
where $\varepsilon$ is a covariantly constant anticommuting 11 dimensional 
spinor. The other symmetry is the local $\kappa$ symmetry
\eqn\kapsym{\eqalign{
\delta_{\kappa} \Theta & = 2 P_+ \kappa (\sigma) \cr
\delta_{\kappa} X^M & = 2i \bar{\Theta} \Gamma^M P_+  \kappa (\sigma)
}}
where $\kappa$ is an 11 dimensional spinor and $P_{\pm}$ are projection 
operators
\eqn\projop{ 
P_{\pm} = \half ( 1 \pm {1 \over 3!} \varepsilon^{\alpha \beta \gamma} 
\partial_{\alpha} X^M \partial_{\beta} X^N \partial_{\gamma} X^P \Gamma_{MNP} )
}
obeying
\eqn\projops{\eqalign{
& P_{\pm}^2  = P_{\pm} \cr
& P_+ P_-  = P_- P_+ = 0 \cr
& P_+ + P_- =1
}}
Here 
$\Gamma_{MNP}= {1 \over 3!}(\Gamma_M \Gamma_N \Gamma_P 
\pm {\rm 5 permutations.})$

For a bosonic membrane configuration ($\Theta =0$) the condition for unbroken
SUSY is that $\delta_{\varepsilon} \Theta =0$. From \glosym\ this seems to be 
impossible. However configurations that differ by a $\kappa$ transformation 
are to be identified so a supersymmetry generated by $\varepsilon$ is unbroken
if there exists a function $\kappa(\sigma)$ such that
\eqn\ubrudt{
\delta_{\varepsilon} \Theta + \delta_{\kappa} \Theta = \varepsilon +2 P_+ \kappa (\sigma) =0
}
Since $P_+$ is a projection operator this equation for $\kappa(\sigma)$ has a 
solution if and only if $\varepsilon = P_+ \varepsilon$ or equivalently
\eqn\pminus{
P_- \varepsilon = 0
}
Since we are only interested in static configurations we take 
$X^0(\sigma^0 , \sigma^1, \sigma^2 ) = \sigma^0$ and $X^M$, $M=1,..,10$ to be 
a function of only $\sigma^1$ and $\sigma^2$. For this configuration the condition on
$\varepsilon$ becomes 
\eqn\condeps{
(1- \half \varepsilon^{\alpha \beta} \partial_{\alpha} X^M \partial_{\beta} X^N \Gamma_{MN} \Gamma_0 )
 \varepsilon =0
}
with $\alpha ,\beta =1,2 $, $\; M,N=1, \ldots, 10$ and $\Gamma_{MN}=\half 
(\Gamma_M \Gamma_N - \Gamma_N \Gamma_M)$.
Now the 10 dimensional space is a \Kahler\ manifold with metric 
$g_{i \bar{j}}$. In many cases there are several choices of complex structure 
which makes the manifold \Kahler. We will return to this point later. 
For now let us assume no more than the space being \Kahler. 
We have gamma matrices $\Gamma_i , \Gamma_{\bar{i}}$,
$i=1,..5$ which obey
\eqn\gammaiden{\eqalign{
\{ \Gamma_i ,\Gamma_{\bar{j}} \} & = 2 g_{i \bar{j}} \cr
\{ \Gamma_i ,\Gamma_j \} & =  \{ \Gamma_{\bar i} ,\Gamma_{\bar j} \} = 0  \cr
 \Gamma_{\bar i} & = ( \Gamma_i )^{\dagger}
}}
The 32 complex dimensional representation of this Clifford algebra can be built
 from a highest weight vector $\varepsilon$ satisfying
\eqn\heighv{
\Gamma_i \;  \vare = 0  \;\;\;\;\;\;\; i=1,..,5
}
by applying the lowering operators $\Gamma_{\bar i}$. This $\vare$ is not 
Majorana. In Calabi-Yau compactifications we know that this $\vare$, 
together with others, is unbroken by the compactification. Thus it makes 
sense to ask which membrane configurations preserve this $\vare$. We have to 
solve the problem of which configurations, $X^M(\sigma^1,\sigma^2)$, solve 
\condeps\ for this $\vare$. First we are free to change coordinate system on 
the worldvolume. It is well known, from string theory for instance, 
that we can  choose coordinates, at least locally, such that the metric 
$h_{\alpha \beta}$ 
is on the form
\eqn\smetric{
h_{\alpha \beta} = g(\sigma^1 , \sigma^2) \delta_{\alpha \beta} 
}
Here we are just displaying the spatial part of the metric. Define a complex 
structure on the worldvolume by $u = \sigma_1 + i \sigma_2 $. The statement 
is now that the supersymmetry generated by $\vare$ is preserved if and only if
the configuration is a holomorphic map, i.e. $X^i(u)$ is holomorphic. 
To prove this we should 
prove that \condeps\ is true if and only if $X^i(u)$ is holomorphic.

First we note that \condeps\ is an equation on each point of the membrane. 
In a given point we can always, for simplicity, choose coordinates in 
spacetime such that $g_{i \bar j}= \half  \delta_{ij}$. 
Writing $z_k = x_k + i y_k$ the condition \heighv\ becomes 
$\Gamma_{x_k} \Gamma_{y_k} \vare = i \vare$. This implies
\eqn\kiral{
\Gamma_{x_1} \Gamma_{y_1} \ldots \Gamma_{x_5} \Gamma_{y_5} \vare = i \vare
}
Working in conventions with $\Gamma_0 \ldots \Gamma_{10} = -1$
this implies $\Gamma_0 \vare = i \vare $. \condeps\ now becomes
\eqn\condepsto{
i \half {\vare}^{\alpha \beta} \partial_{\alpha} X^M  \partial_{\beta} X^N 
\Gamma_{MN} \vare = \vare
}
Using \heighv\ this splits into several equations. From the coefficient of 
$\Gamma_{\bar i} \Gamma_{\bar j}$ we get
\eqn\lignen{
\partial_1 X^{\bar i} \partial_2 X^{\bar j} = \partial_1 X^{\bar j} \partial_2 X^{\bar i}
}
From the coefficient of the unit matrix we get 
\eqn\lignto{
\half i {1 \over{h_{11}}} ( \partial_1 X^{i} \partial_2 X^{\bar i} - \partial_2 X^{i} \partial_1 X^{\bar i}) 
 = 1
}
We also get equations from \fmetric\ and \smetric\
\eqn\ligntre{
h_{11}=h_{22} = \partial_1 X^{i} \partial_1 X^{\bar i} = \partial_2 X^{i} \partial_2 X^{\bar i}
}
\eqn\lignfire{
0= h_{12}= \half ( \partial_1 X^{i} \partial_2 X^{\bar i} + \partial_1 X^{\bar i} \partial_2 X^{i})
}
Combining \lignto, \ligntre\ and \lignfire, we get
\eqn\lignfem{
i \partial_1 X^{i} \partial_2 X^{\bar i} 
= \partial_1 X^{i} \partial_1 X^{\bar i}
}
or
\eqn\lignseks{
i \partial_2 X^{\bar j} \partial_1 X^{i} \partial_2 X^{\bar i} 
= \partial_1 X^{i} \partial_1 X^{\bar i} \partial_2 X^{\bar j}
}
valid for all $j$. Using \lignen\ we get
\eqn\lignsekshalv{
i \partial_2 X^{\bar j} \partial_1 X^{i} \partial_2 X^{\bar i} 
= \partial_1 X^{i} \partial_1 X^{\bar j} \partial_2 X^{\bar i}
}
or
\eqn\lignsyv{
\partial_1 X^{\bar j} - i \partial_2 X^{\bar j} = 0
}
Here we used that 
$ \partial_1 X^{i} \partial_2 X^{\bar i} \neq 0 $ 
which follows from \lignto. \lignsyv\ exactly tells us that 
$X^{\bar j}$ is antiholomorphic or equivalently $X^j$ is holomorphic
in $u=\sigma^1 + i \sigma^2$

The $\vare$ which satisfied the equation is not Majorana. However $P_{-}$ is 
a real operator, so the equation $P_- \vare =0$ is solved by the real and 
imaginary part separately. Alternatively the complex conjugate of $\vare$ 
also satisfies $P_- \vare^* =0 $. $\vare^*$ is the highest weight vector 
with respect to the complex conjugate complex structure. 
This complex structure gives the manifold the opposite orientation.

The result of the discussion above is as follows. Consider M-theory on 
${\bf R}^{1,0} \times K^{10}$, where the first factor is time and $K^{10}$ is 
an oriented Riemannian manifold which admits a complex structure compatible 
with the orientation and which makes it \Kahler. Let $\vare$ be a covariantly
constant spinor satisfying
\eqn\gameps{
\Gamma_i \; \vare = 0 \;\;\;\;\;\; i = 1, \ldots , 5
}
Furthermore consider a membrane with its time direction along time and its 
spatial part embedded in $M^{10}$. Then this configuration preserves the SUSY 
given by the real and imaginary part of $\vare$ if and only if the spatial 
part of the membrane is a holomorphic curve in $M^{10}$.

This prescription allows us to determine the unbroken SUSY of a membrane 
configuration. Below we will consider various special cases. Let us first 
consider ${\bf R}^{1,10}$ with standard metric, $\eta_{ab}$. 
Define $z_1 = x_1 + i x_2$, $\ldots$, $z_5 = x_9 +i x_{10}$. 
The Dirac spinors constitute a 32 dimensional complex vectorspace with a basis 
given by 
\eqn\dspin{
\vare = (\vare_1, \ldots , \vare_5) \;\;\;\;\;\;\; \vare_i = \pm1 
}
where each $\vare_i$ is 1 or -1 depending on whether $\Gamma_i \vare = 0$ or
$\Gamma_{\bar i} \vare = 0$ respectively. Obviously $(1,1,1,1,1)$ is highest 
weight vector for the complex structure $(z_1,z_2,z_3,z_4,z_5)$ and 
$(1,-1,-1,1,-1)$, for example, is the highest weight vector for the complex 
structure $(z_1,\bar{z_2} ,\bar{z_3} ,z_4,\bar{z_5})$ and so on. 
A membrane configuration is holomorphic in both $z$ and $\bar{z}$ 
if and only if $z$ is constant along the membrane. 
We can now consider several cases.

\medskip

\noindent
{\bf 1.} The planar membrane, where $z_5$ is holomorphic and 
$z_1, \ldots ,z_4$ are constant. We see that the preserved 
supersymmetries are given by
\eqn\flatmem{
(\vare_1 ,\vare_2,\vare_3,\vare_4,1) \;\;\;\;\; 
{\rm with} \  \  \vare_1 \vare_2 \vare_3  \vare_4 =1
}
where the last condition came from the requirement that the complex 
structure is compatible with the orientation. Remembering the complex 
conjugate of $\vare$ we see that 16 supersymmetries are unbroken and they 
satisfy
\eqn\flatmemto{
\Gamma_1 \ldots \Gamma_8 \vare = \vare
}
This is, of course, the expected result for the planar membrane.

\medskip

\noindent
{\bf 2.} Consider now a membrane holomorphically embedded in $z_4,z_5$ and 
constant in $z_1,z_2,z_3$. Furthermore assume the embedding is non-degenerate, 
i.e., the membrane is not embedded in a 2-plane inside $z_4,z_5$. 
The preserved SUSYs are
\eqn\curvemem{
(\vare_1 ,\vare_2,\vare_3,1,1) \;\;\;\;\; 
{\rm with} \  \ \vare_1 \vare_2 \vare_3  =1
}
Remembering the complex conjugate
\eqn\curvememto{
(\vare_1 ,\vare_2,\vare_3,-1,-1) \;\;\;\;\; 
{\rm with} \  \ \vare_1 \vare_2 \vare_3  =-1
}
we see that there are 8 unbroken supersymmetries and they satisfy
\eqn\curvememtre{\eqalign{
\Gamma_1 \Gamma_2 \Gamma_3 \Gamma_4 \Gamma_5 \Gamma_6 \Gamma_7 \Gamma_8 \vare & = \vare  \cr
\Gamma_1 \Gamma_2 \Gamma_3 \Gamma_4 \Gamma_5 \Gamma_6 \Gamma_9 \Gamma_{10} \vare & = \vare
}}
This is the same as two perpendicular planar membranes. This last case is 
exactly the case we are interested in, namely the membrane embedded in a 
four real dimensional plane.

\medskip

\noindent
{\bf 3.} It is obvious to extend to the case where the membrane is embedded in 
6,8 and 10 real dimensions. We preserve respectively 4,2 and 2 SUSYs and the 
unbroken SUSY is the same as for intersecting membranes.

Finally let us briefly discuss the case of a Calabi-Yau three fold with 
the membrane wrapped around a holomorphic 2-cycle. Let $z_3,z_4,z_5$ be 
holomorphic coordinates for the Calabi-Yau. The unbroken SUSY by the 
compactification is the real part of $(\vare_1 ,\vare_2,1,1,1)$ and 
$(\vare_1 ,\vare_2,-1,-1,-1)$, i.e. there are 8 unbroken supersymmetries. 
With the membrane the unbroken supersymmetries are
\eqn\cymem{\eqalign{
(\vare_1 ,\vare_2,1,1,1) \;\;\;\;\; {\rm with}\  \ \vare_1 \vare_2 & =1  \cr
(\vare_1 ,\vare_2,-1,-1,-1) \;\;\;\;\; {\rm with}\  \ \vare_1 \vare_2 &  =-1
}}
,i.e., there are 4 unbroken supersymmetries. In the next sections we will 
use the results of this section to construct BPS configurations
of the membrane.

\newsec{$(p,q)$ Strings from Membranes}

Type IIB string theory (IIB) with complexified string coupling 
$\tau = \tau_1 + i \tau_2$ is obtained by compactifying M-theory 
on a torus with complex structure $\tau$. Consider 
M-theory on ${\bf R}^{1,8} \times T^2$ parametrized by 
$(X^0, X^1, \cdots, X^9, X^{10})$ with the identifications
\eqn\identix{
(X^9,X^{10}) \sim (X^9 + 2\pi R, X^{10})
\sim (X^9 + 2\pi R \tau_1, X^{10} + 2\pi R \tau_2).
}
For finite R this describes IIB on a circle. In the limit $R\goto 0$, we 
recover IIB in ten dimensions.

The $(p,q)$ strings in IIB are easily described in this setting. They 
are simply the membranes with one circle wrapped on the torus along the 
$(p,q)$ homology cycle. Specifically, a $(p,q)$ string oriented along the 
$X^1$-axis, say, is described by a membrane embedded as follows.
\eqn\singlestring{\eqalign{
& X^1 = s,\  \ X^9 = 2\pi R t ( p\tau_1 + q ),\  \
X^{10} = 2\pi R t (p \tau_2) \cr
& s\in{\bf R},\  \ t\in [0,1]
.
}}
The overall sign of $(p,q)$ depends on a choice of the orientation of both
the string and the membrane.

We are interested in the 3-string junction, which is located in a plane. 
In M-theory this junction is described by a single membrane. This membrane
has a nontrivial behavior in the plane of the junction and in the torus. From 
the previous section, we know that a BPS configuration is given
by choosing a complex structure in these four dimensions and embedding the 
membrane along a holomorphic curve. The position of the membrane is fixed
in the other six spatial dimensions. We also saw that this BPS configuration 
preserves $1/4$ of the SUSY.

Let the junction lie in the $(X^1,X^2)$ plane and choose the complex structure.
\eqn\complex{
z^1 = X^1 + i X^9,\  \ z^2 = X^2 + i X^{10}.
}
The identifications defining the torus are now
\eqn\identiz{
(z^1, z^2) \sim (z^1 + i 2\pi R, z^2) 
\sim (z^1 + i 2\pi R \tau_1, z^2 +i 2\pi R \tau_2).
}
Define
\eqn\uandv{
u=\exp\left({z^1\over R} - {\tau_1\over \tau_2}{z^2\over R}\right),\  \
v=\exp\left({z^2\over \tau_2 R}\right).
}
We see that $(u,v)\in ({\bf C} - \{0\})^2$ are single valued and constitute a 
global coordinate system on our two complex dimensional manifold 
${\bf R}^2 \times T^2$. 

What is the equation for a $(p,q)$ string? To be a $(p,q)$ string, 
the membrane has to be oriented along the $(p,q)$ homology cycle on the 
$T^2$. In other words, the membrane embedding obeys
\eqn\pqstringx{
p \tau_2 X^9 = (p\tau_1 + q) X^{10}+{\rm const.} ,\   \ {\rm or}
\  \ \Im(p\tau_2 z^1 - (p\tau_1 + q)z^2) = {\rm const.}
}
Since the embedding has to be holomorphic, the equation must be
\eqn\pqstringuv{
p\tau_2 z^1 - (p\tau_1 + q)z^2 = {\rm const}.
}
The real part of this equation shows that the $(p,q)$ string has a 
fixed orientation in the $(X^1,X^2)$ plane given by
\eqn\xyplane{
p \tau_2 X^1 = (p\tau_1 + q) X^2 + {\rm const.}
}
In other words, the $(p,q)$ string is directed along the unit vector
\eqn\unitvector{
{1\over \sqrt{ (p\tau_1 + q)^2 + (p\tau_2)^2}} (p\tau_1 + q, p\tau_2).
}
in the $(X^1,X^2)$ plane. Specifically, the $(0,1)$ string (D-string) is 
oriented along the $X^1$ axis. We recover the observation in 
\refs{\mukhi, \sen} that the type of the string is correlated with 
its orientation. We can write the equation \pqstringuv\ for a single $(p,q)$ string in 
terms of $u$ and $v$ 
as
\eqn\pqinuv{
u^p v^{-q} = \lambda,
}
where the nonzero complex constant $\lambda$ determines the position of
the string on ${\bf R}^2\times T^2$. 

From this discussion we also see that all BPS saturated string network are 
planar. This is because the internal torus is 2 dimensional. Fixing the 
type of a string is the same as fixing the behaviour of the membrane in the internal torus. 
Since this is the imaginary part of an equation the direction of the string in 
space is then fixed and must lie in the 2 plane which together with the torus 
makes a 2 complex dimensional space.  

Before we find the equation describing a 3-string junction, 
let us digress to discuss the metric on ${\bf R}^2\times T^2$ 
since we need it later in order to calculate 
the area of the membrane configuration. The metric is
\eqn\metric{
ds^2 = dx_1^2 + dx_2^2 + dx_9^2 + dx_{10}^2
} 
In our complex structure $(z^1, z^2)$ the manifold is \Kahler\ with the
\Kahler\ form equal to 
\eqn\kahlerz{
\omega = {i\over 2} (dz^1 \wedge d\bz^1 + dz^2 \wedge d\bz^2 )
}
In terms of $(u,v)$, $\omega$ is
\eqn\kahleruv{
\omega = {i\over 2}R^2 
\left\{ 
{du \wedge d\bar{u}\over |u|^2} + |\tau|^2 {dv \wedge d\bar{v}\over |v|^2}
+\tau_1 \left( {du\wedge d\bar{v} \over u\bar{v}}
+ {dv \wedge d\bar{u} \over v\bar{u}} \right)
\right\}
}
%

\newsec{3-String Junction}

In this section, we will derive the equation describing the membrane 
corresponding to a 3-string junction in IIB. We will start with the simplest 
case which is a junction with a $(1,0)$, $(0,1)$ and a $(-1,-1)$ string. 
This is the same case as was considered in \mukhi.
Later we will present the general case.
The membrane configuration is given by a holomorphic curve which is the 
zero locus of a holomorphic function,
\eqn\curve{
f(u,v) = 0.
}

To find the function $f(u,v)$, we use the fact that it should look like one of
the three strings away from the junction. We expect the vertex to be smoothed
out. We want the $(0,1)$ string to be recovered for large $X^1$ and 
$X^2 \approx $const. This means that for $u$ fixed at a very large value
we want exactly one solution in $v$. Similarly, the $(1,0)$ string is recovered
for large $v$, so for fixed large $v$, we want exactly one solution in $u$.
The most general form of the equation with these two properties is
\eqn\almost{ 
uv + au + bv + c = 0,\  \ {\rm or} \  \ 
(u - \lambda_1)(v- \lambda_2 ) = \lambda_3
}
with $\lambda_1, \lambda_2, \lambda_3$ complex constants. If $\lambda_3 =0$, 
the curve is reducible and the equation describes two intersecting planar 
membranes. This is not what we want, so $\lambda_3 \neq 0$.

Let us analyze the curve described by \almost\ for $u\goto\infty$. 
In this limit, $v\goto \lambda_2$ which, according to \pqinuv, describes a 
$(0,1)$ string extended along the $X^1$-direction. We also see the 
geometrical significance of the parameter $\lambda_2$. $\lambda_2$ gives 
the location of the $(0,1)$ string in ${\bf R}^2\times T^2$.
For $v\goto\infty$, we similarly get $u=\lambda_1$, which is the $(1,0)$ 
string at a position given by $\lambda_1$.

What about the $(-1,-1)$ string? We expect to see this for $u,v\goto 0$. 
This is only possible if $\lambda_3 = \lambda_1 \lambda_2$. Now the equation
becomes
\eqn\solution{
uv - \lambda_1 v - \lambda_2 u = 0.
}
For $u,v\goto 0$, the first term is negligible and the equation becomes
\eqn\uvtozero{
uv^{-1} = - {\lambda_1 \over \lambda_2}
}
Combining with \pqinuv, we conclude that this describes a $(1,1)$ or 
$(-1,-1)$ string. Since it is oriented towards small $u$ and $v$
 we see from \uandv\ and \unitvector\ that it is a $(-1,-1)$ string. 

The solution \solution\ thus has the property that there are exactly 3 ways
 to go to 
infinity where the solution becomes respectively a $(0,1)$, $(1,0)$ and
$(-1,-1)$ string. This is enough to conclude that \solution\ is the M-theory
description of the 3-string junction.

There is one immediate advantage in this description of the 3-string junction.
Eq. \solution\ is easily seen to describe a smooth curve. In other words,
the junction has no singularities associated with it. This is in contrast 
to the SYM theory on the D-string \mukhi\ which is not well-suited to capture
 the nature of the vertex.

Strictly speaking the above analysis is only valid when the torus is large
compared the 11 dimensional Planck scale since we have used a low energy 
action to describe the membrane. The 10 dimensional type IIB theory is only 
recovered in the limit where the torus shrinks to zero size. However usual
BPS arguments show that the state will remain BPS for all values of the area 
of the torus, thereby proving that the three string junction in IIB is a 
BPS state.

Having explained the junction with a $(1,0)$,$(0,1)$ and $(-1,-1)$ string 
in detail we will just state the equation for the general junction. 
Consider the junction made of a $(p_1 ,q_1 )$,$(p_2,q_2)$ and
 $(-p_1-p_2 ,-q_1 - q_2 )$ string. Assume furthermore that 
$p_1 q_2 - p_2 q_1 > 0$. If this is not the case we can always relable the 
indices. The equation for this junction is 
\eqn\genjunction{
\lambda_1 u^{-p_1}v^{q_1} + \lambda_2 u^{p_2} v^{-q_2} =1
}
where $\lambda_1$ and $\lambda_2$ are two non-zero complex numbers specifying 
the position of the junction. To see that this equation really describes 
the junction one checks, as before, that there are 3 ways of going to 
infinity and that the equation here describes a string. By setting 
$(p_1 , q_1)= (1,0)$ and $(p_2, q_2) = (0,1)$ we recover the special
case \solution.
This proves that any three string junction obeying charge conservation
\chargecons\ exists. These configurations are not all S-duality transforms 
of each other. Many of them are genuinely different. 
 
\newsec{String Network}

Having set up the formalism, it is easy to generate other curves
and see what they correspond to in the IIB picture. Clearly, there are
very many possibilities. We will not attempt to classify all possible 
configurations. Instead, we will give two examples of string networks to 
illustrate the idea. For simplicity, we will only use $(0,\pm1), (\pm1,0), 
(\pm1,\pm1)$ strings to build the networks. The figures will be drawn for
$\tau = i$.

\medskip
\fig{String Networks. The dotted lines show how the shape of the
lattice (b) changes with the parameter e.}{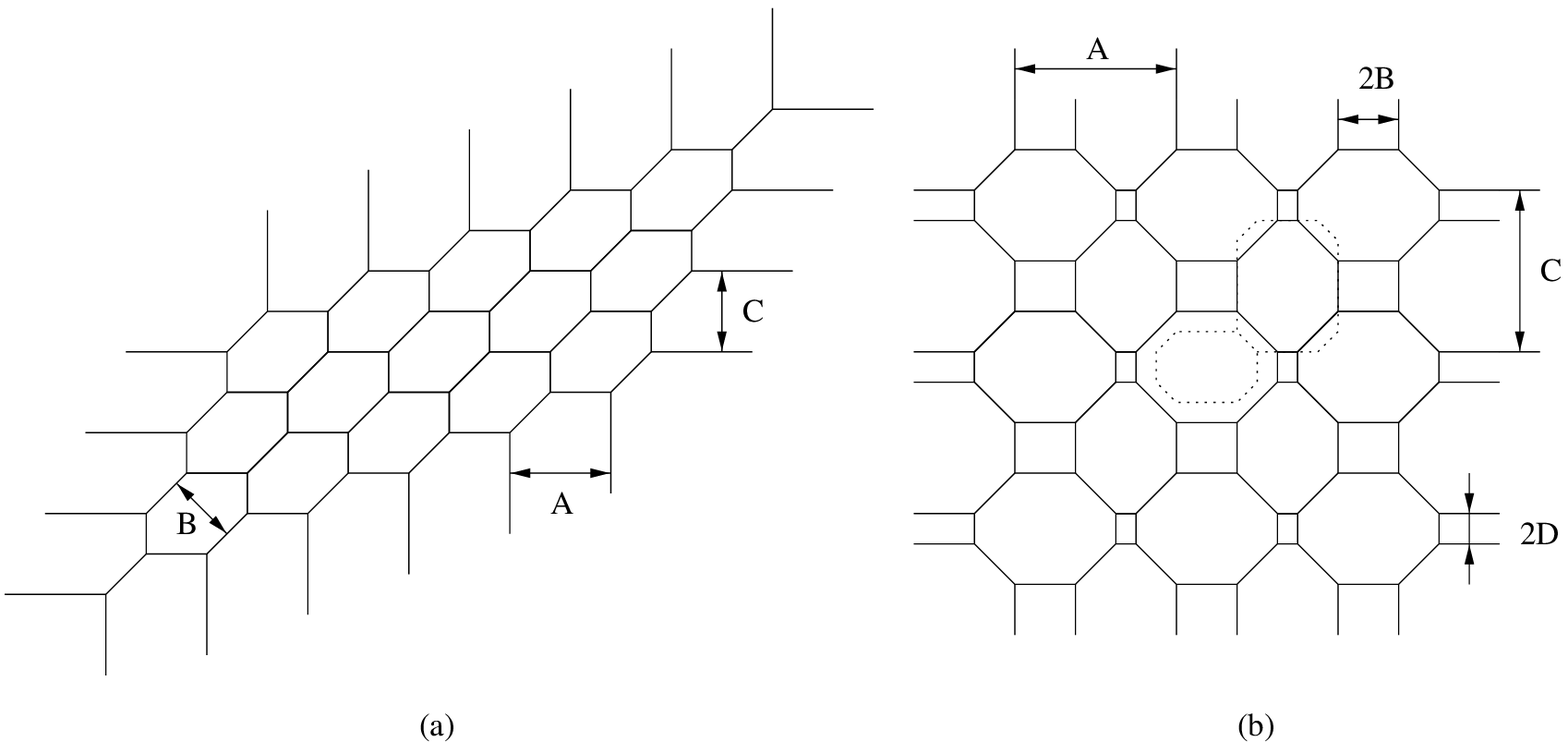}{400pt}
\figlabel\netone
\medskip

The simplest network is drawn in Fig.\netone (a). The angles are fixed by the
types of strings in the network, but we are free to change three lengths 
$A,B,C$. 
We can write down the equation for the network for arbitrary number of unit 
cells. Infinite lattice is obtained by taking an appropriate limit.
For, $(2j-1)\times(2k-1)$ hexagonal cells, the equation is
\eqn\onegeneral{
\sum_{l=-k}^{k}\left\{ v^l P_l(b,c) u^{-j} 
\prod_{m=1}^{2j}(u-b^l a^{2m-2j-1})\right\} = 0,
}
where $P_l(c,d)$ is defined by
\eqn\plcd{
\sum_{l=-k}^{k}v^{l+k}P_l(b,c) = \prod_{m=1}^{2k}(v-b^j c^{2m-2k-1}).
}
The parameters in the equation are related to the lengths by
\eqn\lengths{
A = 2R\ln a,\  \ C = 2R\ln c,\  \ B = \sqrt{2} R\ln (ac/b).
}
To see that \onegeneral\ indeed describes the network, let us look at the 
case $j=k=1$ in some detail. The equation becomes
\eqn\onespecial{
(u-ba)(u-ba^{-1})v^2 - b(c + c^{-1})(u-a)(u-a^{-1})v 
+ b^2 (u-b^{-1}a)(u-b^{-1}a^{-1}) = 0.
}
First, note that the asymptotics of the equation correctly produce 
the eight external lines. The polynomials in $u$ multiplying each $v^l$, 
$(0\le l \le 2k)$ have the factors that specify the positions of the 
internal lines.

In the second example depicted in Fig. \netone (b), there are five 
independent lengths we can change. $A,B,C$ and $D$ are shown in the figure.
The fifth parameter, $e$, represents the freedom to deform the shape of the
lattice without changing the asymptotics. 
The equation for this network is as follows.
\eqn\twogeneral{
\sum_{l=0}^{2k-1}(v^l + v^{-l})Q_l(c,d)
S\left(u;a,{b+e \over 2} +(-1)^{2k-1-l}{b-e\over 2}\right),
}
where $S(u;a,t)$ and $Q_l(c,d)$ are defined by
\eqn\ess{\eqalign{
S(u;a,t) \equiv u^{-2j+1} \prod_{m=-j+1}^{j-1}(u-a^m t)(u-a^m t^{-1}), \cr
\sum_{l=0}^{2k-1}(v^l +v^{-l})Q_l(c,d) 
= v^{-2k+1}\prod_{m=-k+1}^{k-1}(v-c^m d)(v- c^m d^{-1}).
}}
The parameters $a,b,c,d$ are related to the physical parameters by
\eqn\lengthtwo{
A= R\ln a,\  \ B = R\ln b,\  \ C = R\ln c, \  \ D = R\ln d.
}
Apart from the correct asymptotics, note that \twogeneral\ factorizes to give
intersecting fundamental and D-strings when $e=b$.

\newsec{Energy of the 3-String Junction}

The Born-Infeld (BI) description of the fundamental string ending on 
D-p-branes shows that such configurations have no binding energy 
\refs{\calmal, \gibbons} for $p\ge3$. 
For $p = 2$, the D-p-branes do not become flat asymptotically 
and the binding energy is not well-defined.
For $p=1$, the BI theory becomes sigular at the vertex and
is not appropriate to calculate the binding energy.
In this section, we calculate the energy of the 3-string junction
using the M-theory description derived in the previous section.
The binding energy is shown to be zero as expected.

In M-theory, the energy of a 2-brane in its ground state is simply the area
of the 2-brane multiplied by the 2-brane tension 
$T_{M2} = {1 \over (2\pi)^2 l_{11}^3}$.\foot{
Our convention for $l_{11}$ is that $16\pi G_{11} = (2\pi)^8 l_{11}^9$, 
where $G_{11}$ is the 11-dimensional Newton's constant.}
The area is obtained by integrating the \Kahler\ form of the 2-brane, which
is the pull-back of the \Kahler\ form of the ${\bf R}^2\times T^2$ given by
\kahleruv. 

To be definite, let us work with the simplest junction given by \solution.
Set also $\lambda_1 = \lambda_2 = 1$. 
This amounts to locating the junction at the origin in the $(X^1,X^2)$ plane 
as well as fixing the position in the internal torus.
If we choose $u$ as the coordinate on the 2-brane, the 
\Kahler\ form becomes
\eqn\areaform{
\omega = {i\over 2} R^2 \;
\Re\left\{ {1\over \bar{u}(u-1)} + |\tau|^2 {1\over |u-1|^2 u}
+ |\tau +1|^2 {1\over |u|^2(1-u)}\right\} du\wedge d\bar{u}.
}
Note that each of the three terms gives a divergent integral at 
$u = \infty, 1, 0$, respectively. The divergence comes from the infinite 
length of the three strings. We will introduce cutoffs, 
$\Lambda_{(p,q)} \gg 1$, for each $(p,q)$ string. 
Specifically, the integration will be limited to the regions 
$ |u| \le \Lambda_{(0,1)}, |u-1| \ge \Lambda_{(1,0)}^{-1}, 
|u| \ge \Lambda_{(-1,-1)}^{-1}$ for the three terms, 
respectively. The integral is easy to evaluate and the result is 
\eqn\area{ 
A = 2\pi R^2 \sum_{p,q} |p\tau+q|^2 \ln \Lambda_{(p,q)}
}

In order to understand this result in IIB, recall that the 
tension of a $(p,q)$ string is the length of the $(p,q)$ homology cycle 
of the torus times the tension of the membrane (M2),
\eqn\tension{
T_{(p,q)} = 2\pi R |p\tau +q| T_{M2}.
}
We also need to know the relation between the cutoff in the $u$-plane and 
the length of each string. From \uandv\ and \pqinuv, it is clear that
$\ln \Lambda_{(0,1)}$ is the length of the D-string divided by $R$. 
In the same way, one can show that the length of the $(p,q)$ string 
is given by
\eqn\length{
L_{(p,q)} = R |p\tau +q| \ln \Lambda_{(p,q)} + O (\Lambda^{-1}). 
}
The $O(\Lambda^{-1})$ correction becomes negligible in the uncompactified 
IIB limit $(R\goto 0)$.
Combining \area, \tension\ and \length, we obtain
\eqn\energy{
E = T_{M2} A = \sum_{p,q} L_{(p,q)} T_{(p,q)}.
}
We see that the energy is precisely the sum of the energy of the three strings
and there is no binding energy.

\newsec{Discussions}

We have constructed the M-theory realization of the 3-string junction and 
string networks. All the properties of the static configurations are easily 
derived in this formulation. This approach is also suitable for analyzing
the dynamics of these systems\rey. The propagation of wave through the 
junction could be studied by solving the equation of motion (e.o.m.) that 
follows from the membrane action in the static background. For example, 
when the fluctuation is transverse to both $(X^1,X^2)$ and $(X^9,X^{10})$ 
planes, the linearized approximation to the e.o.m. is simply a Helmholtz's 
equation on the (curved) membrane. 

Strictly speaking, the membrane description cannot be trusted when the 
compactification torus is smaller than the 11 dimensional Planck scale.
However, it may worth comparing the result from the membrane action with 
other approaches such as the Born-Infeld theory \rey, or boundary conformal
field theory.

\bigskip
\centerline{\bf Acknowledgements}
We are grateful to Y. K. E. Cheung and L. Thorlacius for discussions. 
The work of MK was supported by the Danish Research Academy.
The work of SL was supported in part by a DOE grant DE-FG02-91ER40671.

\listrefs

\end